\documentclass[11pt]{article}
\usepackage{epsf}
\usepackage{amssymb}
\begin{document}
\title{\bfseries Dependence of critical level statistics \\ on the sample 
shape}
\author{H. Potempa and L. Schweitzer\\
{\itshape Physikalisch-Technische Bundesanstalt, Bundesallee 100,}\\ 
{\itshape D-38116 Braunschweig, Germany}}
\date{}
\maketitle
\begin{abstract}
The level-spacing distribution of consecutive energy eigenvalues is calculated 
numerically at the metal insulator transition for 3d systems with different 
cuboid shapes. 
It is found that the scale independent critical $P_c(s)$ changes as a 
function of the aspect ratio of the samples while the critical disorder 
$W_c/V=16.4$ remains the same. 
We use our data to test whether an expression for the small-$s$ behaviour 
of the level statistics proposed by Kravtsov and Mirlin for the metallic 
regime is applicable also at the critical point.
For this reason, a shape dependent dimensionless critical conductance $g_c$ 
has been extracted from the small-$s$ behaviour of the critical level 
statistics. 
Our result for a cubic sample, $g_c=0.112\pm 0.005$, is in good agreement 
with a value obtained previously from calculations using the Kubo-formula.  
\end{abstract}

\section{Introduction}
Energy eigenvalue correlations provide general tools for the statistical
description of disordered materials. Among them, the nearest neighbour 
level spacing distribution $P(s)$ represents one of the 
simplest statistics. Here, $s=|E_{i+1}-E_i|/\Delta$ is the energy difference
of consecutive eigenvalues $E_i$ divided by the mean level spacing $\Delta$.
Nevertheless, knowing $P(s)$, it is possible to tell whether a given system 
exhibits metallic or insulating behaviour at temperature $T=0$\,K.

Considerable attention has recently focused on the special case of critical
statistics which were found at the metal insulator transition in disordered
3d \cite{Sea93,HS93a,HS94a,AKL94,AKL95,ZK95,BSZK96,CKL96,ZK97} 
and 2d \cite{SZ95,Eva95,OO95,BS97} systems.
In all cases, the numerically obtained critical level statistics were noticed 
to be independent of the system size $L$. This is different in the
metallic and insulating regimes where a size dependence has been observed. 
In the limit $L\to\infty$, the nearest neighbour spacing distribution follows
the Poissonian decay, $P(s)=\exp(-s)$, for disorder strength $W$ larger than 
the critical disorder $W_c$. The metallic side, $W/W_c < 1$, is well described 
by random matrix theory (RMT) \cite{Wig57,Dys62,Meh91,Efe83}. 
As in the metallic phase, the critical $P_c(s)$ depends 
on the symmetry of the Hamiltonian describing the system under consideration. 
In addition, and in contrast to the universal RMT description for the metallic 
phase, the critical $P_c(s)$ depends also on the spatial dimension. 
The respective forms of $P_c(s)$ are known only from numerical studies.

In Ref.\,\cite{BSZK96} using the Anderson model, the application of an 
Aharonov-Bohm-flux was shown not only to change the universality class of 
the level statistics due to  breaking of time reversal symmetry. 
It has also been demonstrated that, depending on the strength of the flux, 
there exists a set of scale independent critical spacing distributions. 
These are associated with a 
continuous crossover from orthogonal to unitary symmetry \cite{BSZK96,Mon97}. 
It is well known that the corresponding Hamilton matrix can be transformed 
such that the AB-flux is completely absorbed into the boundary conditions. 
So the question arises, whether or not $P_c(s)$ does in general depend on the 
boundary conditions applied. Recently, it has been shown \cite{BMP98} that 
$P_c(s)$ is in fact sensitive to a change of the periodic boundary conditions, 
which were always applied previously, to Dirichlet boundary conditions in one,
two and three directions.

\begin{figure}
\epsfxsize12.5cm\epsfbox{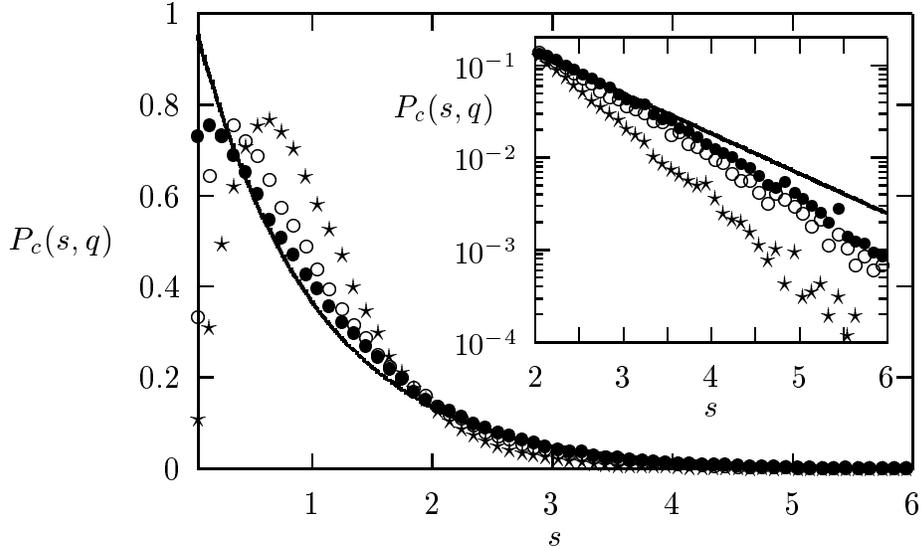}
\caption[]{\label{fig1}The critical level spacing distribution $P_c(s,q)$ 
versus the level spacing $s$ for aspect ratios $q=1$ ($\star$, 
$L_0=L_z=30\,a$), $q=1/7$ ($\circ$, $L_0=56\,a, L_z=8\,a$), and $q=20$ 
($\bullet$, $L_0=10\,a, L_z=200\,a$). The solid line is the Poissonian decay 
$\exp(-s)$. The insert displays the semi-logarithmic plot of the large-$s$ 
behaviour.}  
\end{figure}

In this paper we show that $P_c(s)$ is sensitive to the shape of the sample, 
too. We present results of a numerical investigation on various finite size 
cuboids with square base $L_0^2$ and height $L_z$, characterized by the 
aspect ratio $q=L_z/L_0$. Assuming that a formula for the small-$s$ behaviour
of $P(s)$, proposed by Kravtsov and Mirlin \cite{KM94} for the metallic 
regime, may approximately be used at the metal-insulator transition,
a dimensionless critical conductance $g_c$ can be extracted from the 
small-$s$ behaviour of the calculated $P_c(s)$. We find that $g_c$ also
depends on the sample shape and that the value obtained for the cubic system
is in close agreement with a result obtained by a different method.

\section{Model}
We investigate the conventional 3d Anderson model which is described by the
Hamiltonian
\begin{equation}
H=\sum_m \varepsilon_m^{} c_m^\dagger c_m^{} + V \sum_{<m\ne n>} 
(c_m^\dagger c_n^{} + c_n^\dagger c_m^{}).
\end{equation}
The creation, $c_m^\dagger$, and annihilation, $c_m^{}$, operators act on the
states of non-interacting electrons at the sites $\{m, n\}$ of a 3d simple 
cubic lattice. 
The disorder energies $\varepsilon_m$ are chosen to be a set of independent 
random numbers distributed in 
the interval $[-W/2, W/2]$ with probability $1/W$, where $W$ denotes the 
disorder strength measured in units of $V$. We consider only transfer between
nearest neighbour sites. Periodic boundary conditions are applied in all 
directions. The size of the different systems investigated are described by 
a quadratic base $L_0^2$ and a height $L_z$ ranging from 8 to 200 lattice 
constants $a$. The shape of the system is conveniently denoted by the 
dimensionless aspect ratio $q=L_z/L_0$. For all shapes, a size independent
metal insulator transition is found at a critical disorder of $W_c/V=16.4$.

The eigenvalues used in our study have been taken from an interval 
$[-0.5\,V, 0.5\,V]$ around the band centre. 
They are obtained by direct diagonalization using a Lanczos algorithm. 
For each aspect ratio, several system sizes are computed with a large number 
of different realizations of the disorder potentials so that for each 
particular case the number of eigenvalues exceeded $10^5$.  

\section{Results and Discussion}
In Fig.~\ref{fig1} the size independent critical level spacing distribution 
$P_c(s,q)$ is shown for different aspect ratios: a flat quasi 2d sample 
($q=1/7$), a cube ($q=1$), and a long narrow bar ($q=20$). Deviations of the 
aspect ratio from $q=1$, result in critical spacing distributions that 
seem to approach the Poissonian decay, $P(s)=\exp(-s)$, as $q \to 0$ 
(quasi 2d sample) or $q \to \infty$ (quasi 1d sample).

\begin{figure}
\epsfxsize12cm\epsfbox{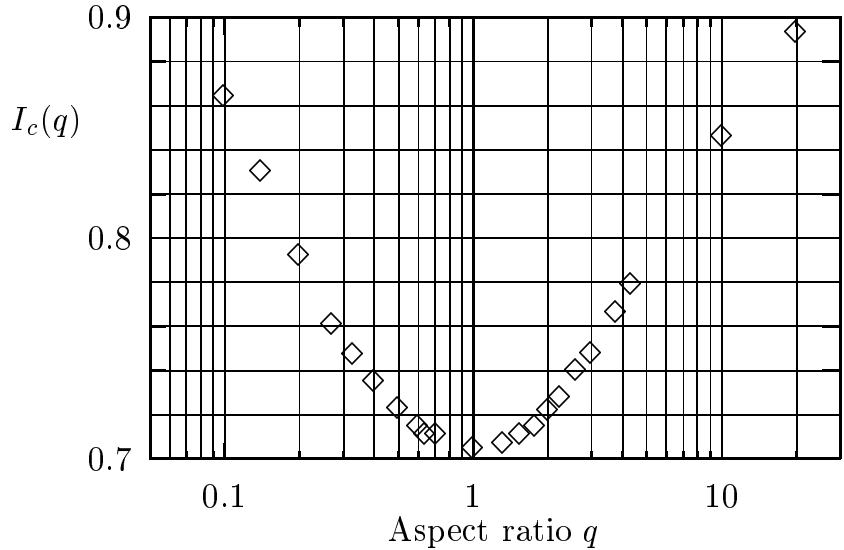}
\caption[]{\label{fig2}The second moment of the critical level spacings, 
$I_c(q)$, versus the aspect ratio $q=L_z/L_0$.The corresponding system sizes 
are in the range from ($L_0/a=80, L_z/a=8$) to ($L_0/a=10, L_z/a=200$).}  
\end{figure}

To give a more quantitative description, we have calculated the second moment 
$I_c(q)$ of the critical level spacings 
\begin{equation}
I_c(q)=1/2 <s^2>=1/2 \int_0^\infty s^2 P_c(s,q)\,ds
\end{equation}
which is shown in Fig.~\ref{fig2} as a function of the aspect ratio 
$q=L_z/L_0$. The cube ($q=1$) represents the minimum value of the second 
moment, $I_c(1)=0.705 \pm 0.005$, whereas $I_c(q)\to 1$ for aspect ratios that 
strongly deviate from the cube. Our value of $I_c(q=1)$ is in accord with 
the result of Ref.~\cite{ZK95}. From the symmetric appearance of 
the semi-logarithmic plot in Fig.~\ref{fig2} it becomes apparent that 
$I_c(q)\to I_c(1/q)$ when crossing $q=1$. 

At present, there exists no complete analytical theory which describes the
critical energy level correlations.
To compare our results, we therefore have to put up with a formula proposed 
recently \cite{KM94} for the metallic regime. 
The two-level correlation function is defined as
\begin{equation}
R(s)=\frac{1}{<\rho>^2}<\rho(E)\rho(E+\omega)>, 
\end{equation}
where $\rho(E)$ is the density of states at energy $E$, 
$s=\omega/\Delta$, and $<...>$ denotes averaging over realizations
of the disorder potential. In the metallic regime and for small $s$, 
$R(s)$ becomes \cite{KM94} 
\begin{equation}
R(s)=\frac{\pi^2}{6}(1+\frac{3b}{\pi^6g^2})s.
\label{KM}
\end{equation}
This expression contains corrections to the usual Wigner-Dyson form 
in the region $\omega\sim\Delta << E_c$, where $E_c$ is the Thouless energy. 
Eq.~(\ref{KM}) depends on the dimensionless conductance $g>>1$, the spatial 
dimensions, the boundary conditions and also on the shape of the system via 
the diffusive modes \cite{KM94}. For cuboids $L_0^2\times L_z$ and periodic 
boundary conditions one gets
\begin{equation}
b=\frac{9/16}{(2+q^2)^2}
\sum_{{n_x, n_y, n_z=-\infty\atop n_x^2+n_y^2+n_z^2\ne 0}}^\infty
\frac{1}{(n_x^2+n_y^2+n_z^2q^{-2})^2}.
\end{equation} 
At the critical point, where $g=g_c\lesssim 1$, the result of the expansion 
made in Ref.~\cite{KM94} is expected to give only qualitatively correct 
behaviour. 
In order to test this expectation, we tentatively assume here that 
Eq.~(\ref{KM}) actually holds at the metal insulator transition.
Therewith, we are able to extract a shape dependence of the size independent 
critical conductance $g_c$ from the small-$s$ part of the calculated 
$P_c(s,q)$, because for small $s$, 
$R(s)=\sum_k P(k,s)+\delta (s)\simeq P(0,s)$,
where $P(k,s)$ is the probability density to find exactly $k$ eigenvalues 
within the interval $s$. 

\begin{figure}
\epsfxsize12cm\epsfbox{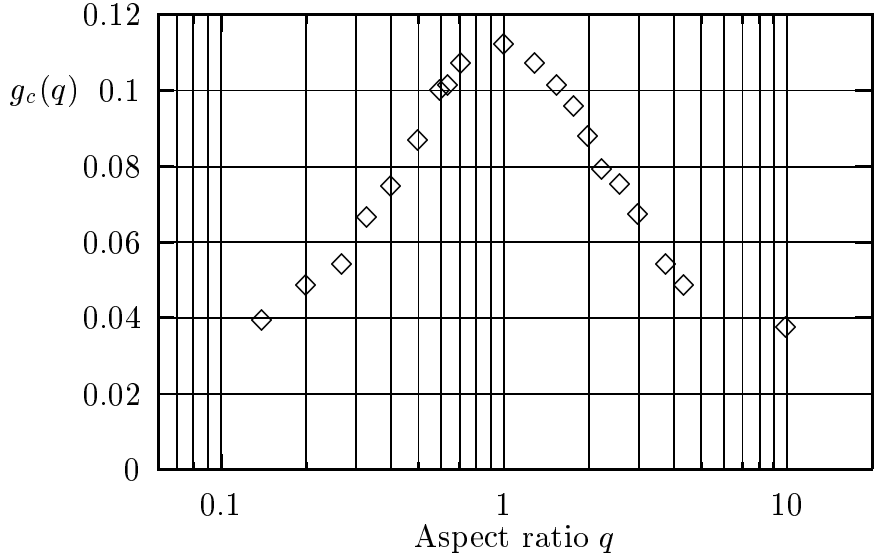}
\caption[]{\label{fig3}The shape dependent critical conductance $g_c(q)$ 
as a function of the aspect ratio $q$.}  
\end{figure}

The critical conductance as a function of the sample shape is shown in 
Fig.~\ref{fig3}. Here, the values for $q=0.1$ and $q=20$ have been omitted,
because it was not possible to accurately extract the small-$s$ gradient 
from the data. 
Again, the already noticed symmetry in replacing $q\to 1/q$ when crossing
$q=1$ becomes evident.
 
Our result for the cubic system, $g_c(1)=0.112\pm 0.005$, 
agrees very well with the value $g_c=0.10\pm 0.01$ obtained from the numerical 
calculation of the Kubo-conductivity \cite{LS94}. 
Since the two methods for determining the critical conductance are completely 
different, the perfect agreement of the numerical values is remarkable. 
If this accordance is not mere accidental, it has to be understood 
why the result of the expansion in Ref.~\cite{KM94} holds also quantitatively
at the critical point. 
For Dirichlet boundary conditions, we obtain a numerical value 
$g_c=0.079\pm0.002$ for the cubic system. 

In mesoscopic systems, the decay of $g_c(q)$ accompanied with deviations 
from the cubic shape may explain some of the different values (0.03-0.2) 
\cite{KM87} reported for measurements on various samples. However,
the question remains to be answered, whether or not the expression for
the small-$s$ behaviour proposed by Kravtsov and Mirlin \cite{KM94} for
the metallic regime is really applicable at the metal insulator transition.  

In conclusion, a dependence of the scale independent critical level 
statistics on the shape of the samples has been detected at the Anderson 
transition in 3d systems. 
Using a formula which has been proposed for the metallic regime \cite{KM94} 
to fit our data obtained at the critical point, we find a shape dependence of 
the scale independent conductance $g_c$.
Provided that this method of extracting the conductance is justified also at 
the critical point, the various experimentally obtained critical conductances 
reported in the literature can possibly be attributed to different shapes of 
the samples investigated.

\section*{Acknowledgments}
We would like to thank G.~Montambaux and I.~Kh.~Zharekeshev for useful
discussions.


\begin{thebibliography}{10}
\itemsep-2pt
\bibitem{Sea93}
B.~I. Shklovskii {\it et~al.}, Phys.\ Rev.\ B {\bf 47},  11487  (1993).

\bibitem{HS93a}
E. Hofstetter and M. Schreiber, Phys. Rev. B {\bf 48},  16979  (1993).

\bibitem{HS94a}
E. Hofstetter and M. Schreiber, Phys. Rev. Lett. {\bf 73},  3137  (1994).

\bibitem{AKL94}
A.~G. Aronov, V.~E. Kravtsov, and I.~V. Lerner, JETP Lett. {\bf 59},  39
  (1994).

\bibitem{AKL95}
A.~G. Aronov, V.~E. Kravtsov, and I.~V. Lerner, Phys.\ Rev.\ Lett. {\bf 74},
  1174  (1995).

\bibitem{ZK95}
I.~K. Zharekeshev and B. Kramer, Jap. Journ. Appl. Phys. {\bf 34},  4361
  (1995).

\bibitem{BSZK96}
M. Batsch, L. Schweitzer, I. {Kh. Zharekeshev}, and B. Kramer, Phys.\ Rev.\
  Lett. {\bf 77},  1552  (1996).

\bibitem{CKL96}
J.~T. Chalker, V.~E. Kravtsov, and I.~V. Lerner, JETP Lett. {\bf 64},  386
  (1996).

\bibitem{ZK97}
I. {Kh. Zharekeshev} and B. Kramer, Phys.\ Rev.\ Lett. {\bf 79},  717  (1997).

\bibitem{SZ95}
L. Schweitzer and I. {Kh. Zharekeshev}, J. Phys.:\ Condens.\ Matter {\bf 7},
  L377  (1995).

\bibitem{Eva95}
S.~N. Evangelou, Phys.\ Rev.\ Lett. {\bf 75},  2550  (1995).

\bibitem{OO95}
T. Ohtsuki and Y. Ono, J. Phys.\ Soc.\ Jpn. {\bf 64},  4088  (1995).

\bibitem{BS97}
M. Batsch and L. Schweitzer,  in {\em High Magnetic Fields in the Physics of
  Semiconductors II: Proceedings of the International Conference, W\"u{}rzburg
  1996}, edited by G. Landwehr and W. Ossau (World Scientific Publishers Co.,
  Singapore, 1997), pp.\ 47--50.

\bibitem{Wig57}
E.~P. Wigner, Ann. Math. {\bf 65},  203  (1957).

\bibitem{Dys62}
F.~J. Dyson, Journal of Mathematical Physics {\bf 3},  140  (1962).

\bibitem{Meh91}
M.~L. Mehta, {\em Random Matrices}, 2nd ed. (Academic Press, San Diego, 1991).

\bibitem{Efe83}
K.~B. Efetov, Adv. Phys. {\bf 32},  53  (1983).

\bibitem{Mon97}
G. Montambaux, Phys.\ Rev.\ B {\bf 55},  12833  (1997).

\bibitem{BMP98}
D. Braun, G. Montambaux, and M. Pascaud, cond-mat/9712256  (1997).

\bibitem{KM94}
V.~E. Kravtsov and A.~D. Mirlin, JETP-Lett. {\bf 60},  645  (1994).

\bibitem{LS94}
P. Lambrianides and H.~B. Shore, Phys. Rev. B {\bf 50},  7268  (1994).

\bibitem{KM87}
M. Kaveh and N.~F. Mott, Phil. Mag. B {\bf 55},  9  (1987).

\end{thebibliography}
\end{document}